\documentclass[amsmath,amssymb,11pt,superscriptaddress,reprint, preprintnumbers, notitlepage,aps,twocolumn,nofootinbib]{revtex4-1}
\pdfoutput=1 
\usepackage[utf8]{inputenc}
\usepackage[english]{babel}
\usepackage{amsmath}
\usepackage{graphicx}
\usepackage{dcolumn}
\usepackage{pbox}
\usepackage{amssymb}
\usepackage{epsfig}
\usepackage[dvipsnames]{xcolor}
\usepackage[normalem]{ulem}
\usepackage{slashed}
\usepackage{amssymb}
\usepackage{ mathrsfs }
\usepackage{color}
\usepackage[font=small]{caption}
\usepackage[font=small]{subcaption}
\usepackage{url}
\usepackage[makeroom]{cancel}
\usepackage{tikz}
\usetikzlibrary{positioning} 
\usetikzlibrary{arrows}
\usepackage{enumitem} 
\usepackage{soul}
\usepackage{comment}
\allowdisplaybreaks

\newcommand\scalemath[2]{\scalebox{#1}{\mbox{\ensuremath{\displaystyle #2}}}}

\definecolor{MyDarkBlue}{rgb}{0.1, 0.1, 0.8} 
\definecolor{SBlue}{rgb}{0.2, 0.4, 0.7} 
\definecolor{MyLightBlue}{rgb}{0.22,0.51,0.9}
\definecolor{MyGreen}{rgb}{0.0, 0.5, 0.0}
\definecolor{BrickRed}{rgb}{0.8, 0.25, 0.33}
\RequirePackage{hyperref}
\hypersetup{colorlinks, citecolor=SBlue,linkcolor=MyDarkBlue, urlcolor=PineGreen}

\makeatletter
\renewcommand\@makecaption[2]{%
  \par
  \vskip\abovecaptionskip
  \begingroup
  
   \small\rmfamily
    \begingroup
     \samepage
     \flushing
     \let\footnote\@footnotemark@gobble
     \@make@capt@title{#1}{#2}\par
    \endgroup
  \endgroup
  \vskip\belowcaptionskip
}
\makeatother

\makeatletter 
    
\renewcommand\onecolumngrid{
\do@columngrid{one}{\@ne}%
\def\set@footnotewidth{\onecolumngrid}
\def\footnoterule{\kern-6pt\hrule width 1.5in\kern6pt}%
}

\renewcommand\twocolumngrid{
        \def\footnoterule{
        \dimen@\skip\footins\divide\dimen@\thr@@
        \kern-\dimen@\hrule width.5in\kern\dimen@}
        \do@columngrid{mlt}{\tw@}
}%

\makeatother    

\DeclareUnicodeCharacter{2212}{-}
\begin{document}
\title{\vspace{1cm}\Large 
Metastable Cosmic Strings and Gravitational Waves \\from Flavour Symmetry Breaking 
}

\author{\bf Stefan Antusch}
\email[E-mail:]{stefan.antusch@unibas.ch}
\author{\bf Kevin Hinze}
\email[E-mail:]{kevin.hinze@unibas.ch}
\affiliation{Department of Physics, University of Basel, Klingelbergstrasse\ 82, CH-4056 Basel, Switzerland}
\author{\bf Shaikh Saad}
\email[E-mail:]{shaikh.saad@okstate.edu}
\affiliation{Jožef Stefan Institute, Jamova 39, P.\ O.\ Box 3000, SI-1001 Ljubljana, Slovenia}

\begin{abstract}
Metastable cosmic strings (MSCSs) are among the best-fitting explanations of the 2023 pulsar timing array (PTA) signal for gravitational waves at nanohertz frequencies. We propose the novel possibility that a network of MSCSs generating this signal originates from the multi-step spontaneous breaking of a gauged flavour symmetry. As a specific example, we construct a model of $SU(2)$ flavour symmetry in the context of $SU(5)$ grand unification, where the $SU(2)$ acts exclusively on the first two generations of the matter 10-plet, such that it is ``right for leptons'' and allows for large lepton mixing. The model explains the mass hierarchies of the Standard Model fermions, and predicts the string scale of the MSCSs in a range compatible with the 2023 PTA signal. Cosmic inflation is associated with the latter step of (two-step) family symmetry breaking, and the phase transition ending inflation generates the cosmic string network. 
\end{abstract}

\maketitle

\section{Introduction}

In summer 2023, pulsar timing arrays (PTAs)~\cite{NANOGrav:2023gor,Antoniadis:2023ott,Reardon:2023gzh,Xu:2023wog} 
found strong evidence of Hellings-Downs angular correlation~\cite{Hellings:1983fr} in pulsar time-of-arrival perturbations, providing a first clear signal of a stochastic gravitational wave background (SGWB) at nanohertz frequencies. 
Remarkably, metastable cosmic strings (MSCSs) arising from gauge groups are among the best-fitting explanations~\cite{NANOGrav:2023hvm} of this signal, which has sparked interest within the particle physics community, see Refs.~\cite{Antusch:2023zjk,Buchmuller:2023aus,Fu:2023mdu,
Lazarides:2023rqf,Ahmed:2023rky,Afzal:2023cyp,Maji:2023fhv,Ahmed:2023pjl,Afzal:2023kqs,King:2023wkm,Roshan:2024qnv,Ahmed:2024iyd,Antusch:2024ypp,Maji:2024pll,Antusch:2024nqg,Datta:2024bqp,Maji:2024cwv,Pallis:2024joc,Ahmad:2025dds} for works on metastable strings, and Refs.~\cite{Lazarides:2023ksx,Yamada:2023thl,Servant:2023mwt,
Lazarides:2023bjd,Pallis:2024mip} for other works on unstable cosmic strings. 

In the above mentioned studies the origin of MSCSs has been linked to particle physics extensions where the gauge group of the Standard Model (SM) is embedded into a larger (e.g.\ unified) group, which is then spontaneously broken to the SM in multiple steps, generating first monopoles~\cite{Kibble:1976sj,Preskill:1979zi,Vilenkin:2000jqa} and then cosmic strings~\cite{Kibble:1976sj,Vilenkin:2000jqa}. If the two symmetry breaking scales are close, and if certain conditions are satisfied ~\cite{Vilenkin:1982hm, Preskill:1992ck}, monopole-antimonopole pairs can form along the strings through quantum tunneling, causing the strings to decay and resulting in the metastability of the cosmic string network. MSCSs can generate a GW spectrum across a wide range of frequencies, with a characteristic drop, as observed by the PTAs, when the decay happens.
However, in addition to extensions of the gauge symmetry of the SM towards unification of forces, there exists another class of symmetry extensions of the SM, namely ``\textit{flavour symmetries}'', which act ``horizontally'' between different generations.

Flavour symmetries are postulated to address the so-called ``\textit{flavour puzzle}'' of the SM, which becomes apparent when considering the SM flavour sector:  
The SM flavour sector contains in total fourteen unknown parameters, which include the six quark masses, three charged lepton masses, and four quark mixing angles. Experimental determination of these parameters reveals a strong hierarchical pattern in the masses and mixings of the charged fermions. While neutrinos are massless in the SM, observations from various experiments have confirmed neutrino oscillations, implying nine (seven, if neutrinos are Dirac particles) additional parameters, including three neutrino masses, three mixing angles, and three CP-violating phases. Unlike the charged fermion sector, the mixing angles in the neutrino sector are large. The lack of a fundamental understanding of the strong hierarchical structure of the charged fermions -- spanning approximately six orders of magnitude -- and the apparently anarchical structure of the neutrino sector is one of the major challenges in particle physics, referred to as the flavour puzzle.

In this paper, we propose the novel possibility that a network of MSCSs generating the 2023 PTA signal originates from the spontaneous breaking of a gauged ``\textit{flavour symmetry}''. 
In Sec.~\ref{sec:01},  we outline the specifications for generating metastable cosmic strings and discuss how they can emerge, for example, from an \(SU(2)\) flavour symmetry. In Sec.~\ref{sec:02}, we construct a realisation of a flavour model with \(SU(2)\) flavour symmetry in the context of \(SU(5)\) grand unification. Finally, we conclude in Sec.~\ref{sec:03}.

\section{MSCS\MakeLowercase{s} and Flavour Symmetry}\label{sec:01}
When the flavour symmetry group $G_F$ and its breaking dynamics satisfies the following criteria, it can give rise to a MSCS network: To start with, the theory has to give rise to cosmic strings as well as monopoles, both originating from the same parent group (here $G_F$). We consider here the multi-step breaking of $G_F$ to subgroups, for example $G_F \to G_F' \to G_F''$, but of course the breaking may happen in a larger number of steps. 
Monopoles~\cite{Kibble:1976sj,Preskill:1979zi} are generated when, e.g.\ for the breaking $G_F \to G'_F$, the second homotopy group of the quotient space $G_F/G'_F$ is non-trivial, i.e.\ when $\pi_2(G/G^\prime)\neq {1}$, while cosmic strings~\cite{Kibble:1976sj} appear in case of a non-trivial first homotopy group, e.g.\ when $\pi_1(G'_F/G''_F)\neq {1}$.  When for the breaking $G_F \to G_F' \to G_F''$ both above-mentioned conditions are satisfied, but the homotopy groups of $G_F/G''_F$ are all trivial, then none of these topological defects can be stable. The monopoles and cosmic strings form an unstable hybrid topological defect.  

Whether the cosmic strings, generated after the monopoles, are long-lived enough to explain the 2023 PTA observations, depends on further criteria: The first one is that the abundance of monopoles produced from the breaking $G_F \to G_F'$ has to be diminished before cosmic string production, to avoid that the cosmic strings that form at the later breaking $G'_F \to G''_F$ attach to the monopoles and antimonopoles, which would cause a too quick decay of the cosmic string network. 
One way to resolve this is to realise a phase of inflation~\cite{Guth:1980zm, Albrecht:1982wi, Linde:1981mu, Linde:1983gd} between the two stages of symmetry breaking, diluting the monopoles. Nevertheless, pairs of monopoles and antimonopoles can spontaneously nucleate along the strings by quantum tunneling, making the cosmic strings decay.
However, this process is only efficient if these two formation scales are close to each other. If they are well separated, the cosmic string network effectively behaves as a stable one. Therefore, this second condition -- that these scales are of the same order -- guarantees that the lifetime of the hybrid network coincides with the PTA signal. Last but not least, this hybrid network must originate from a gauge group to be capable of explaining the PTA data.

\subsection{SU(2) Flavour Symmetry Example}
To illustrate the above let us consider a non-Abelian\footnote{Abelian groups, cf.~\cite{Blasi:2024vew}, lead to stable strings and cannot explain the recent PTA data.} gauged $SU(2)_F$ flavour symmetry, which is a popular route towards addressing why the first two families of charged fermions are lighter than the third (cf.~\cite{Barbieri:1995uv, Barbieri:1996ae, Barbieri:1996ww, Barbieri:1999pe, Masiero:2001cc, Linster:2018avp, Barbieri:1997tu, Carone:1997qg,  Roberts:2001zy, Raby:2003ay, Dudas:2013pja, Falkowski:2015zwa, Feruglio:2019ybq, Barbieri:2019zdz, Linster:2020fww, Greljo:2023bix,Antusch:2023shi,Greljo:2024zrj} for models using $U(2)$ or $SU(2)$  as flavour symmetries). 
When the first two generations of fermions are doublets of $SU(2)_F$, then their Yukawa couplings (i.e.\ finally their masses) can only emerge from an effective operators, explaining their smaller values, while the masses of the third generation stem from an unsuppressed renormalisable operator.  
The hierarchies in the masses of the first two generations of fermions can then further emerge e.g.\ from a two-step breaking of $G_F$ by the vacuum expectation values (VEVs) of two types of so-called flavons, i.e.\ scalar fields that break the family symmetry. 

To generate a network of MSCSs explaining the PTA results, we propose that the flavour   symmetry breaking proceeds as follows:
\begin{align}
&SU(2)_F\xrightarrow[]{\langle \Delta\rangle}   \underbrace{U(1)_F\xrightarrow[]{\langle \phi\rangle+\langle \overline\phi\rangle}}_{\mathrm{inflation}} \text{{\em nothing}},   \label{eq:flavour-symmetr-breaking}
\end{align}
where $\Delta$ is a \( SU(2)_F \) triplet flavon  and $\phi + \overline{\phi}$ is a pair of \( SU(2)_F \) doublet flavon  superfields.\footnote{Since the string and monopole scales turn out to be very high, close to the grand unification scale, we consider in our example the case of a supersymmetric theory, to address the hierarchy problem.} 
The first breaking pattern generates topologically stable monopoles~\cite{Kibble:1976sj,Preskill:1979zi,Vilenkin:2000jqa}, whereas the second breaking produces stable cosmic strings~\cite{Kibble:1976sj,Vilenkin:2000jqa}. But since the breaking $SU(2)_F \to \text{{\em nothing}}$ does not give rise to any topological defects, after the breaking is completed both must be unstable. 
Following the discussion in the previous section, a stage of inflation may be realised before the second stage of breaking happens, which gives rise to cosmic strings. 
This implies that inflation is now embedded in the flavour sector of the theory. Attractive candidates are for instance supersymmetric (SUSY) Hybrid inflation~\cite{Linde:1993cn,linde1991axions,Dvali:1994ms}, where the ``waterfall'' (2nd order) phase transition is associated with the breaking of $U(1)_F$ part of the family symmetry, or Tribrid inflation~\cite{Antusch:2004hd,Antusch:2009vg,Antusch:2010va,Antusch:2024qpb}, again ended by the $U(1)_F$ flavour symmetry breaking phase transition, but with the inflaton identified as a matter field (e.g.\ a right-handed sneutrino). 
Finally, much later, monopole-antimonopole pairs forming along the strings through quantum tunneling cause the strings to decay and result in a MSCS network~\cite{Vilenkin:1982hm}.

One remarkable property of MSCSs is that in addition to explaining the 2023 PTA observations, which features a pronounced drop of the GW spectrum at low (nanoherz) frequencies, it predicts a rather flat GW spectrum at larger frequencies, that may span over many orders of magnitudes in frequency. This means that the MSCS explanation of the 2023 PTA signal can be confirmed (or refuted) by various planned future GW observatories, such as 
Laser Interferometer Space Antenna (LISA)~\cite{Audley:2017drz}, Einstein Telescope (ET)~\cite{Sathyaprakash:2012jk}, and Cosmic Explorer (CE)~\cite{Evans:2016mbw}. 
When confirmed, this would select very specific scenarios in particle physics and cosmology.  Furthermore, via the induced deviation from standard cosmology (with SM particle degrees of freedom), signs of extensions of the SM (such as supersymmetry) could be found even when the new particles have masses up to about $10^4$ TeV\ \cite{Antusch:2024ypp}.    

A confirmation of the MSCS origin of the SGWB found by PTAs would also require a new perspective on flavour model building, which we aim to initiate with this letter. In addition to resolving the \textit{flavour puzzle}, successful models then also have to generate the SGWB from MSCSs and realise cosmic inflation in deep connection to flavour symmetry breaking. In the next section, we will present a first model of this type, in the framework of supersymmetric SU(5) grand unification.

\section{Realisation in a Flavour GUT Model}\label{sec:02}

We will now construct a first flavour model capable of explaining the 2023 PTA results from flavour symmetry breaking with embedded inflation, as well the flavour structure of charged fermions as well as neutrinos.  
In contrast to many previous models using this flavour symmetry group, and which were facing problems accommodating the observed large mixing in the lepton sector, we make use of the proposal of \cite{Antusch:2023shi} that in order to elegantly explain large lepton mixing one can choose the $SU(2)_F$ to act in a way that is ``right for leptons and left for quarks''. As discussed there, the idea has an attractive realisation in \( SU(5) \) grand unified theory (GUT)~\cite{Pati:1973rp,Pati:1974yy, Georgi:1974sy, Georgi:1974yf, Georgi:1974my, Fritzsch:1974nn}, where the SM fermions are contained in the \( \overline{5}_F \) and \( 10_F \) superfields (considering a SUSY framework). With only the first two generations of \( 10_F^{1,2} \) forming a doublet under the \( SU(2) \) flavour symmetry, which we now refer to as $SU(2)_{10}$, while the rest of these fields, including the right-handed neutrinos, are \( SU(2)_{10} \) singlets, large lepton mixing is generated along with small quark mixing and the observed charged fermion mass hierarchies. The model we construct in this section is also the first one realising this idea.

\subsection{Fields and Charges}

The model consists of the usual \(SU(5)\) chiral superfields, $\overline 5_F^p + 10_F^p$, with $p=1,2,3$ being the generation index. The $\overline 5_F^p$ and $10_F^3$ fields are singlets under the gauged $SU(2)_{10}$ flavour symmetry, whereas the first two generations $10_F^i$ form a doublet (here, $i=1,2$ corresponds to the $SU(2)_{10}$ index). 
Note that the model we construct below, which is based on  $SU(5) \times SU(2)_{10}$ gauge symmetry, is free from gauge anomalies.

At the high  scale (compatibility with proton decay rate requires $\gtrsim 10^{16}$ GeV), the GUT symmetry is spontaneously broken down to the SM by the VEV of an adjoint superfield $24_H$, $\langle 24_H\rangle=v_{24} \,\mathrm{diag}(-1,-1,-1,3/2,3/2)$. The SM gauge group is subsequently broken by the electroweak doublets living in the $5_H+\overline 5_H$ superfields under \(SU(5)\) group. For convenience we denote these fields as follows:  
\begin{align}
&\chi\equiv \overline 5_F,\;\; \psi\equiv 10_F,
\\
&\Phi\equiv 24_H,\;\; H\equiv 5_H,\;\; \overline H\equiv \overline 5_H.
\end{align}

The flavour symmetry breaking pattern, assisted by the VEVs of $\Delta$ and $\phi+\overline \phi$, is shown in Eq.~\eqref{eq:flavour-symmetr-breaking}.  
By using $SU(2)_{10}$ rotations, one can make the triplet VEV to be diagonal. On the other hand, consistency of $D$-term conditions allow the following generic VEV structure for the pair of doublets:
\begin{align} 
    \langle\phi\rangle =
    \begin{pmatrix} v_1 \\ v_2 \end{pmatrix}
    ,\;\;
    \langle\overline\phi\rangle = 
    \begin{pmatrix} v_1 \\ v_2 \end{pmatrix}e^{i\varphi}, \;\; \langle\Delta\rangle=
\begin{pmatrix}
    v_3&0\\0&-v_3
\end{pmatrix}. \label{eq:vev}
\end{align}

\begin{table}[th]
\centering
\footnotesize
\resizebox{0.35\textwidth}{!}{
\begin{tabular}{|c|c|c|c|c|}
\hline
Symbol& $SU(5)$& $SU(2)_{10}$ & $U(1)_g$ & \( \mathbb{Z}_4 \) 
\\ [1ex] \hline\hline
$\psi^i$& $10^i_F$& 2& +1& +2\\  \hline
$\psi^3$& $10^3_F$& 1& 0& 0\\  \hline
$\chi^p$& $\overline 5^p_F$& 1& 0& +2\\  \hline\hline
$\Phi$&$24_H$&  1& 0& +2\\  \hline
$H$&$5_H$&  1& 0& 0\\  \hline
$\overline H$&$\overline 5_H$& 1& 0& 0\\  \hline\hline
$\Delta$&$1$&  3& $-1$& +3\\  \hline
$\phi$&$1$&  2& +1& +2\\  \hline
$\overline\phi$&$1$&  $\overline 2$& $-1$& +2\\  \hline\hline
$S$&$1$&  1& 0& 0\\  \hline\hline
$\nu_R^p$&$1^p$&  1& 0& +2\\  \hline
\end{tabular}
}
\caption{Charge assignments. }
\label{tab:charge}
\end{table}

In the spirit of explaining the desired fermion mass and mixing hierarchies, we additionally utilise a global \( U(1)_g \) symmetry and a discrete \( \mathbb{Z}_4 \)  symmetry. 
Furthermore, we impose a matter parity where all matter fields are odd. 
The quantum numbers of all these multiplets are summarised in Table~\ref{tab:charge}. With this charge assignments, the Yukawa part of the superpotential can be written as
\begin{align}
W_\mathrm{Y}=  W_\mathrm{u} +W_\mathrm{d} +W_\mathrm{\nu},  \label{eq:Ysuperpotential} 
\end{align}
with 
\begin{align}
W_\mathrm{u} &= H \epsilon_5 \bigg\{  
y_1 \psi^3 \psi^3 + \frac{1}{\Lambda^3} y^{\prime\prime}_2 \psi^i \psi^j \epsilon_{ik} (\Delta^2)^k_j \Phi 
\nonumber\\&
+\frac{1}{\Lambda} y_3 \psi^i\psi^3\overline\phi_i
+\frac{1}{\Lambda^2} y_4 \psi^i\psi^j\overline\phi_i\overline\phi_j
\bigg\}, \label{eq:Wu}
\end{align}
\begin{align}
W_\mathrm{d} &=\frac{1}{\Lambda} \overline H \Phi \bigg\{ y^{\prime(\prime\prime),p}_5 \psi^3 \chi^p
+\frac{1}{\Lambda} y^{\prime(\prime\prime),p}_6 \psi^i \chi^p \overline \phi_i 
\bigg\}
\nonumber\\&
+\frac{1}{\Lambda^3} y^p_7 \overline H \psi^i \chi^p \phi^k \epsilon_{kj} (\Delta^2)^j_i, \label{eq:Wd}
\end{align}
and
\begin{align}
W_\mathrm{\nu}= y^{pq}_\nu \chi^p\nu_R^q H + M^{pq}_R \nu_R^p\nu_R^q. \label{eq:neutrino}
\end{align}

As a theory of flavour, all dimensionless Yukawa couplings are expected to be of order unity.
In the above equations, couplings $y^\prime_i$  and $y^{\prime\prime}_i$ correspond to two distinct $SU(5)$ index contractions~\cite{Dorsner:2024seb}. Moreover, $\epsilon_5$ ($\epsilon_{ij}$) denotes the Levi-Civita tensor for $SU(5)$ ($SU(2)_{10}$) group. To avoid clutter, we have suppressed the $SU(5)$ group indices, however,  $SU(2)_{10}$ index contractions are shown explicitly. The neutrino mass is generated via the type-I seesaw mechanism~\cite{Minkowski:1977sc,Yanagida:1979as,Glashow:1979nm,Gell-Mann:1979vob,Mohapatra:1979ia,Schechter:1980gr}, and it is important to emphasise that the absence of flavon VEVs in the neutrino sector is crucial for achieving large mixing angles.

In Eqs.~\eqref{eq:Wu}-\eqref{eq:Wd}, we have assumed a common cutoff scale, $\Lambda$, and further define $\varepsilon_i \equiv v_i/\Lambda$. The derived mass matrices of the charged fermions (written in the $f^c M_f f$ basis) take the following forms:
\begin{align}
&\scalemath{0.9}
{  
M_u=2\langle H\rangle \bigg\{
\begin{pmatrix}
2y_4 \varepsilon_1^2 & 2y_4 \varepsilon_1\varepsilon_2 &y_3\varepsilon_1
\\
2y_4 \varepsilon_1\varepsilon_2&2y_4\varepsilon_2^2&y_3\varepsilon_2  
\\
y_3\varepsilon_1&y_3\varepsilon_2&2y_1
\end{pmatrix}
+  y_2 \varepsilon_3^2\varepsilon_{24}
\begin{pmatrix}
0 & 1 &0
\\
-1&0&0
\\
0&0&0
\end{pmatrix}
\bigg\}
},\label{eq:mu} 
\\
&
\scalemath{0.93}
{ 
M_d= \frac{\langle \overline H\rangle}{\sqrt{2}}
\bigg\{
\varepsilon_{24} \begin{pmatrix}
d_{1}\varepsilon_1 &d_{1}\varepsilon_2& d^\prime_1
\\
d_{2}\varepsilon_1 &d_{2}\varepsilon_2& d^\prime_2
\\
d_{3}\varepsilon_1 &d_{3}\varepsilon_2& d^\prime_3
\end{pmatrix} 
+
\varepsilon_3^2 \begin{pmatrix}
-c_{1}\varepsilon_2 &c_{1}\varepsilon_1& 0
\\
-c_{2}\varepsilon_2 &c_{2}\varepsilon_1& 0
\\
-c_{3}\varepsilon_2 &c_{3}\varepsilon_1& 0
\end{pmatrix}
\bigg\}
},  \label{eq:md}
\\
&
\scalemath{0.93}
{ 
M_e^T= \frac{\langle \overline H\rangle}{\sqrt{2}}
\bigg\{
\varepsilon_{24} \begin{pmatrix}
\hat d_{1}\varepsilon_1 &\hat d_{1}\varepsilon_2& \hat d^\prime_1
\\
\hat d_{2}\varepsilon_1 &\hat d_{2}\varepsilon_2& \hat d^\prime_2
\\
\hat d_{3}\varepsilon_1 &\hat d_{3}\varepsilon_2& \hat d^\prime_3
\end{pmatrix}   
+
\varepsilon_3^2 \begin{pmatrix}
-c_{1}\varepsilon_2 &c_{1}\varepsilon_1& 0
\\
-c_{2}\varepsilon_2 &c_{2}\varepsilon_1& 0
\\
-c_{3}\varepsilon_2 &c_{3}\varepsilon_1& 0
\end{pmatrix}  
\bigg\}
}, \label{eq:me}
\end{align}
where we have defined, 
\begin{align}
\langle H\rangle = v\sin\beta, \;\langle \overline H\rangle = v\cos\beta,\; v=174\mathrm{GeV},    
\end{align}
and the expressions for $d^{(\prime)}_i, \hat d^{(\prime)}_i, c_i$ in terms of the original Yukawa couplings can be found in    Appendix~\ref{sec:fit}.

An excellent fit to the charged fermion masses and mixings with \(  \mathcal{O}(1)\) coefficients can be obtained from Eqs.~\eqref{eq:mu}-\eqref{eq:me}, and a benchmark parameter set is presented in Appendix~\ref{sec:fit}. On the other hand, due to the absence of flavons in the  neutrino sector, cf.\ Eq.~\eqref{eq:neutrino}, the neutrino mass matrix remains anarchic. Therefore, large mixings angles~\cite{Altarelli:2002sg} in the neutrino sector are naturally predicted within this setup.

\subsection{Embedding of Inflation}
The breaking of the \( SU(5) \) symmetry produces GUT scale monopoles. Moreover, the first stage of symmetry breaking in the flavour sector, \( SU(2)_{10} \to U(1)_F \), also generates monopoles. 
As will be described below, we implement inflation after these two steps of symmetry breaking to dilute away these monopoles.\footnote{Additionally, domain walls and global strings, produced during the first two breakings, are also diluted away by inflation. } 
After inflation, the complete breaking of the remaining \( U(1)_F \) flavour symmetry leads to the formation of metastable cosmic strings, which can have intriguing implications for cosmology.

To realise inflation, we may employ e.g.\ supersymmetric Hybrid inflation~\cite{Linde:1993cn,linde1991axions,Dvali:1994ms}, ended by a ``waterfall'' phase transition identified with \( U(1)_F\) breaking. In this Hybrid inflation scenario, the scalar component of a gauge singlet superfield, \(S\), plays the role of the inflaton, while the pair of doublets \( \phi + \overline{\phi} \) act as the waterfall fields, break the corresponding symmetry, here \( U(1)_F \to \text{{\em nothing}}\), and desirably end inflation. SUSY Hybrid inflation can be realised through the following term in the superpotential:
\begin{align}
W_\mathrm{Inflation} \supset \lambda S(\overline{\phi}\phi-m_{\phi}^2) ,  \label{eq:inflation}  
\end{align}
where $\lambda$ is a dimensionless coupling, and $m_{\phi}$ a mass scale.
Note that the combination \(\overline{\phi}\phi\) of the $SU(2)_F$ doublet flavon superfields is allowed by both the global \( U(1)_g \)  and  \( \mathbb{Z}_4 \) symmetries. 
As long as the VEV of \( S \), which generates masses for the $\phi$- and $\overline{\phi}$-fields, is large enough, the latter are stabilised at zero and the scalar potential features a large vacuum energy $V_\mathrm{inf}=\lambda^2 m_{\phi}^4$ that drives inflation.

While \( V_\mathrm{inf} \) is flat in the $S$ direction at tree level within the minimal SUSY Hybrid inflation setup, 
non-canonical K\"ahler contributions as well as quantum corrections and SUSY breaking effects can provide an appropriate curvature of the inflaton potential for successful inflation in the $S$ direction (cf.\ \cite{Senoguz:2004vu,Bastero-Gil:2006zpr,Rehman:2009nq,Nakayama:2010xf,Antusch:2012jc,Buchmuller:2014epa,Schmitz:2018nhb}). Due to this curvature, as the inflaton rolls towards smaller field values, the $\phi$-fields become tachyonic beyond a critical point, $|S_c|= m_{\phi}$, triggering a waterfall transition where they acquire VEVs, breaking $U(1)_F$ spontaneously and ending inflation. 
In our framework, cosmic strings form 
only after inflation and are therefore not diluted.

In the following, we briefly summarize the main results of SUSY Hybrid inflation, for the example of a minimal K\"ahler potential, and taking into account one-loop corrections to the inflaton potential (as reviewed e.g.\  in \cite{Bastero-Gil:2006zpr}). In this case, the spectral index is predicted to be
\begin{align}
    n_s\approx 1- \frac{1}{N_e} \simeq 0.98 , 
\end{align}
where we have in the last step used  \( N_e \simeq 50 \). This value of $n_s$ is slightly larger, but reasonably close to, the currently preferred experimental estimates~\cite{Planck:2018jri,ACT:2025fju,ACT:2025tim}. 
Moreover, the amplitude of the primordial spectrum can be written as
\begin{align}  
\mathcal{P}_\mathcal{R}^{1/2} &\approx  2 \sqrt{\frac{N_e}{3}} \left(\frac{m_{\phi}}{M_p}\right)^2  ,
\end{align}  
where $M_p$ is the reduced Planck mass. By matching it to the experimental  data~\cite{Planck:2018jri}, we obtain \( m_{\phi} \simeq 6 \times 10^{15} \, \text{GeV} \). 
After inflation, $m_{\phi}$ also sets the cosmic string scale $v_\textrm{cs}$. As we will see in the next section, the prediction from this simple realisation of SUSY Hybrid inflation is well compatible with the range for $v_\textrm{cs}$ that we obtain from a fit of the model to the date for fermion masses and mixing, and which in turn matches well with the string scale required to explain the PTA data.

Note that additional contributions from, e.g., a non-canonical K\"ahler potential or the SUSY breaking sector, can further modify the inflaton potential, allowing to more accurately fit the experimentally favoured value of $n_s$, while enlarging the range of $m_{\phi}$ compatible with CMB data. A detailed exploration of these effects, however, lies beyond the scope of the present paper and will be investigated in a future publication.

\subsection{MSCSs and Predicted SGWB}
When the two scales of flavour symmetry breaking are close to each other, as explained in the introduction, the cosmic strings become metastable with a lifetime large enough to explain the 2023 PTA results. More specifically, the decay of this network has the rate per string unit length  given by~\cite{Vilenkin:1982hm, Preskill:1992ck, Monin:2008mp, Leblond:2009fq, Chitose:2023dam}
\begin{align}
\Gamma_d\simeq \frac{\mu}{2\pi} e^{-\pi\kappa},\;\;\; 
\kappa= \frac{m^2}{\mu}\simeq \frac{8\pi}{g^2} \left( \frac{v_\mathrm{m}}{v_\mathrm{cs}} \right)^2,   \label{decay}
\end{align}
where, $v_\mathrm{m}=v_3$  and $v_\mathrm{cs}=(|v_1|^2+|v_2|^2)^{1/2}$ are the VEVs associated with the string and monopole formation scales, respectively. The monopole mass is $m\simeq 4\pi v_\mathrm{m}/g$, with \(g\) being the relevant gauge coupling constant, and the energy per unit length of the string is $\mu\simeq 2\pi v_\mathrm{cs}^2$.

For a metastable cosmic string network, the recent PTA results~\cite{NANOGrav:2023gor,Antoniadis:2023ott,Reardon:2023gzh,Xu:2023wog} hint towards a string tension in the range $G\mu \simeq \left(10^{-8} - 10^{-5}\right)$ and $\kappa^{1/2} \simeq \left(7.7 - 8.3\right)$, displaying a significant correlation between the two variables~\cite{NANOGrav:2023hvm} (where $G$ is Newton's constant and $G\mu\simeq 4.22\times 10^{-38} v^2_\mathrm{cs}$). For such a range of the metastability parameter $\kappa$,  the most stringent bound on the string tension comes from the LIGO-Virgo-KAGRA (LVK)  collaboration~\cite{LIGOScientific:2014pky,VIRGO:2014yos,KAGRA:2018plz}, which implies $G\mu \lesssim 2\times 10^{-7}$~\cite{LIGOScientific:2021nrg,NANOGrav:2023hvm}  at a frequency $\mathcal{O}(20)$ Hz~\cite{KAGRA:2021kbb}.  However, larger values of \( G\mu \) are fully compatible with non-standard cosmology leading to some dilution of the string network, as discussed in Ref.~\cite{Antusch:2024ypp} and references therein.

To determine the GW spectrum from the metastable cosmic string network in our model, we need to identify the relevant symmetry-breaking scales in the flavour sector. For our analysis, we set \( M_\textrm{GUT} = 2.0 \times 10^{16} \, \text{GeV} \), a value motivated by gauge coupling unification with TeV-scale SUSY.  This high GUT scale ensures adequate suppression of gauge-mediated proton decay lifetimes. Furthermore, adequate metastability of the string network implies \( \epsilon_3 \approx \sqrt{|\epsilon_1|^2 + |\epsilon_2|^2} \), which we also employ in the numerical fit presented in Appendix~\ref{sec:fit}. Now, from the charged fermion masses, Eqs.~\eqref{eq:mu}-\eqref{eq:me}, we obtain the following approximate relations
\begin{align}
    &y_u=\sin\beta\frac{\epsilon_3^4\epsilon_{24}^2}{\epsilon_2^2}\frac{C_u^2}{C_c},\;\;\; y_c=4\sin\beta\epsilon_2^2C_c,\label{eq:c1} \\
    &y_{\tau,b}=\frac{\cos\beta}{\sqrt{2}}\epsilon_{24}C_{\tau,b}, \;\;\; y_{\mu,s}=\frac{\cos\beta}{\sqrt{2}}\epsilon_{24}\epsilon_2 C_{\mu,s},
    \\
    &y_{e,d}=\frac{\cos\beta}{\sqrt{2}}(C_{e,d}+ C_1)\epsilon_3^2\epsilon_2. \label{eq:c2}
\end{align}
Here, $C_f\sim \mathcal{O}(1)$ are some combinations of the original Yukawa couplings and their exact forms are not relevant in determining the cosmic string scale. 

\begin{figure}[t!]
    \centering
    \includegraphics[width=1\linewidth]{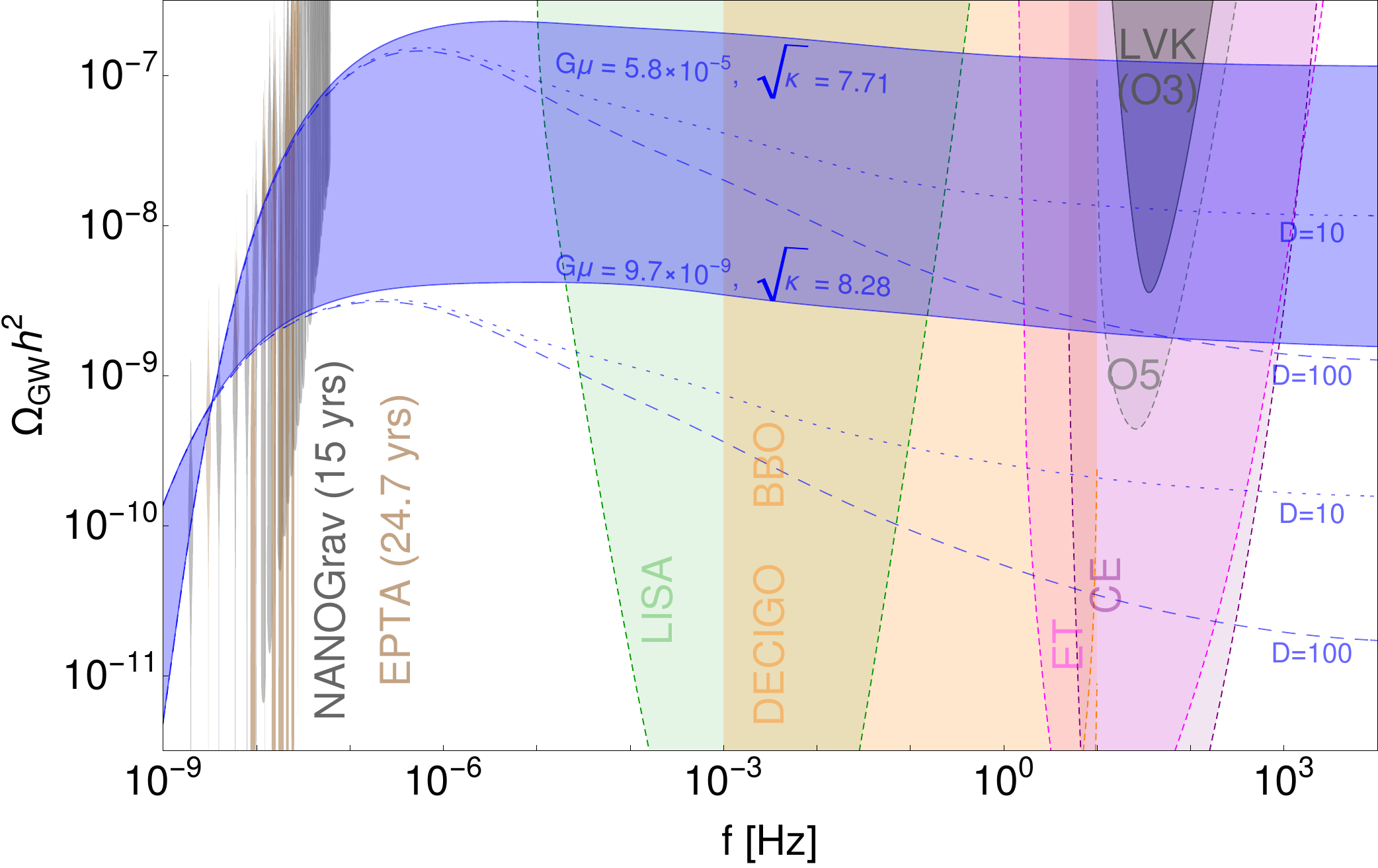}
    \caption{Predicted range for the gravitational wave spectrum of metastable cosmic strings from the proposed GUT flavour model with $SU(2)_{10}$ flavour symmetry. The blue band shows the case of standard cosmology, whereas the dotted (dashed) lines indicate the predicted range for a non-standard cosmology scenario where a late matter-dominated phase leads to  
    a dilution of the GW spectra by a factor $D=10$ ($D=100$). As an example we have taken  $M_\mathrm{SUSY}=3\,$TeV for the scale of sparticle masses.   
    }
    \label{fig:GW}
\end{figure}

Plugging in the GUT scale values for the Yukawa couplings\ \cite{Antusch:2013jca} in Eqs.~\eqref{eq:c1}-\eqref{eq:c2}, and varying the $C_f$ parameters between $0.1$ and $2$, yields
\begin{align}
    &\epsilon_{24}\in[0.027,\,0.805],\;
    \epsilon_3\in[0.013,\,0.062],\\
   \;\;\; &\frac{\epsilon_3}{\epsilon_{24}}\in[0.018,\,2.207],
    \\
    \mathrm{hence,}\;\;\; & v_\textrm{cs}\in[3.6\times 10^{14},\,4.4\times 10^{16}]\ \text{GeV}.\label{eq:range}
\end{align} 
Furthermore, viable points were found for $\tan\beta<24$ (and down to $\tan\beta=5$ which we used as lower prior).
The GW spectrum arising from a metastable cosmic string network for \(v_\textrm{cs}\) in the range given above is depicted in Fig.~\ref{fig:GW}. The details of the computational procedure for this spectrum are relegated to Appendix~\ref{sec:GWspectrum}. The plot illustrates that if the origin of the 2023 PTA result from MSCSs is confirmed, the scenario can be further tested by multiple upcoming GW observatories.
The plot also shows that for a standard cosmological history (blue band in Fig.~\ref{fig:GW}), the LVK O3 constraints $v_\textrm{cs}\lesssim 2\times 10^{15}$\ GeV already exclude a sizable part of the allowed parameter space. 
Excitingly, LVK O5 will probe the entire range, Eq.~\eqref{eq:range}, predicted by the fermion mass hierarchies.  Future GW experiments,  such as  LISA~\cite{Audley:2017drz}, Big Bang Observer (BBO)~\cite{Corbin:2005ny}, DECi hertz Interferometer Gravitational wave Observatory (DECIGO)~\cite{Seto:2001qf}), ET~\cite{Sathyaprakash:2012jk}, and CE~\cite{Evans:2016mbw} will also test this model. 
Furthermore, in Fig.~\ref{fig:GW} we have also included the case that a late matter-dominated phase, typical for certain SUSY scenarios (see e.g.\ \cite{Ellis:1984eq,Endo:2006zj,Nakamura:2006uc,deCarlos:1993wie,Hasenkamp:2010if,Co:2016xti}), leads to some dilution of the GW spectra. The dotted (dashed) lines show the predicted range for a dilution factor $D=10$ ($D=100$). As discussed in \cite{Antusch:2024ypp}, the scenario (including the SUSY scale) is nevertheless testable with a combination of LISA and ET/CE.

\section{Conclusions}\label{sec:03}
In this letter, we proposed that the 2023 pulsar timing array (PTA) signal for gravitational waves (GWs) at nanohertz frequencies is generated by a metastable cosmic string (MSCS) network produced from flavour symmetry breaking. We discussed the requirements for flavour models to explain the PTA results, and constructed a first example model. 
The model is based on a gauged $SU(2)_F$ flavour symmetry in the context of $SU(5)$ grand unification, 
broken spontaneously in two steps, $SU(2)_F\to U(1)_F\to \text{{\em nothing}}$, by VEVs of triplet and doublet flavons, respectively. 
The $SU(2)_F$ acts exclusively on the first two generations of the matter 10-plet, such that it is ``right for leptons'', allowing for large lepton mixing. In addition to explaining the mass hierarchies of the SM fermions, it realises cosmic inflation in association with flavour symmetry breaking and generates a MSCS network with cosmic string scale predicted (for a standard cosmological history) to be in the range \( v_\textrm{cs} \in [3.6 \times 10^{14},\, 4.4 \times 10^{16}] \, \text{GeV} \), compatible with the 2023 PTA signal and fully testable by ongoing and future GW detectors.
Finally, our work introduced a novel direction in flavour model building that not only addresses the renowned \textit{flavour puzzle}, but also deeply connects to the dynamics of cosmic inflation and the GW signatures suggested by PTAs, which might be confirmed in the not too far future.

\section*{Acknowledgments}
SS acknowledges the financial support
from the Slovenian Research Agency (research core funding No.\ P1-0035 and N1-0321).

\appendix
\section{Numerical Fit}\label{sec:fit}
As explained in the main text, hierarchies in the fermion masses are generated dynamically by the VEVs of the flavons and the order in which they contribute to the respective entries in the Yuakwa matrices.  Yukawa couplings, on the other hand, take $\mathcal{O}(1)$ values. 
The redefined Yukawa coefficients appearing in the mass matrices Eqs.~\eqref{eq:mu}-\eqref{eq:me}  are given by
\begin{align}
&d_p=\frac{3}{2}y^{\prime,p}_6 + y^{\prime\prime,p}_6,\;\;\;
d_p^\prime=\frac{3}{2}y^{\prime,p}_5 + y^{\prime\prime,p}_5,\\
&\hat d_p=\frac{3}{2}y^{\prime,p}_6 -\frac{3}{2} y^{\prime\prime,p}_6,\;\;\;\hat d_p^\prime=\frac{3}{2}y^{\prime,p}_5 -\frac{3}{2} y^{\prime\prime,p}_5,\\
&c_p=y^p_7, \;\;\;y_2= \frac{5}{4}y_2^{\prime\prime}.
\end{align}
To show the compatibility of our scenario, here a benchmark fit to the charged fermion sector is presented. The fit is performed at \(M_\mathrm{GUT}\) and we have chosen the SUSY scale to be 3 TeV. To simplify the analysis, SUSY threshold corrections are set to zero. Input values of the observables at the GUT scale are taken from Ref.~\cite{Antusch:2013jca}.   For generality, all Yukawa parameters are taken to be complex, however, we set $\varphi=0$ in Eq.~\eqref{eq:vev}. In our numerical analysis, we allow the absolute values of the Yukawa couplings in the range $(0.1-2)$. The following parameter set reproduces all charged fermion masses and mixings within their  experimentally allowed $1\sigma$ values: 
\begin{align}
&\tan\beta=  7.70,
\\&
\left( \varepsilon_{24},  \varepsilon_1, \varepsilon_2 \right)=
\left( 0.0543, 0.00597e^{0.543 i}, 0.0508e^{0.229 i}\right), 
\\&  
\left( y_4, y_2 \right)=  \left(  0.105e^{-1.551 i}, 0.210e^{-0.0329 i}    \right),
\\&  
\left( y_3, y_1 \right)=  \left(  0.128e^{-1.632 i}, 0.129e^{1.561 i}    \right),
\\&
\left( d_1, d_2, d_3 \right)=  \left( 0.225e^{0.870 i}, 0.2621e^{-2.582 i}, 0.3728e^{1.569 i} \right),  
\\&
\left( d_1^\prime, d_2^\prime, d_3^\prime \right)=  \left(  0.970e^{0.009 i}, 0.508e^{-2.808 i}, 0.136e^{1.550 i} \right), 
\\&
\left( \hat d_1, \hat d_2, \hat d_3 \right)=  \left( 1.758e^{-1.518 i}, 0.237e^{1.490 i}, 0.219e^{-3.008 i} \right),  
\\&
\left( \hat d_1^\prime, \hat d_2^\prime, \hat d_3^\prime \right)=  \left(  0.451e^{-1.810 i}, 0.270e^{0.8389 i}, 1.371e^{-1.651 i} \right), 
\\&
\left( c_1, c_2, c_3\right)=  \left( 1.679e^{0.217 i}, 0.4519e^{-3.044 i}, 1.454e^{-0.0134 i}  \right).
\end{align}
The fitted values at the GUT scale are given by 
\begin{align}
&\left( y_u, y_c, y_t\right)=     \left( 2.485\times 10^{-6}, 1.4239\times 10^{-3}, 0.5121 \right), 
\\
&\left( y_d, y_s, y_b\right)=     \left( 4.976\times 10^{-6}, 9.813\times 10^{-5}, 
5.467\times 10^{-3} \right), 
\\
&\left( y_e, y_\mu, y_\tau\right)=     \left( 1.9919\times 10^{-6}, 4.26\times 10^{-4}, 7.275\times 10^{-3} \right), 
\\
&\left( \theta_{12}^\mathrm{ckm}, \theta_{23}^\mathrm{ckm}, \theta_{13}^\mathrm{ckm}, \delta^\mathrm{ckm}  \right)=     \left( 0.2272, 
0.0396, 0.00346, 1.195 \right) 
\end{align}
where $y_p=y_p^\mathrm{MSSM}\sin\beta$ for $p=u,c,t$ and $y_r=y_r^\mathrm{MSSM}\cos\beta$ for $r=d,s,b,e,\mu,\tau$.

\section{GW spectrum}\label{sec:GWspectrum}
In this appendix we summarize the computational procedure for the GW spectrum of metastable cosmic strings\footnote{We note that the prediction for the SGWS spectrum from MSCS is subject to considerable theoretical uncertainties. For example, the role of string segments is currently under discussion~\cite{Chitose:2025qyt}. In our calculation of the SGWS spectrum, we have not included this possible contribution. A reduction of this uncertainty is highly desirable, and might in the future allow to distinguish different MSCS realisation, such as cases with or without unconfined fluxed~\cite{NANOGrav:2023hvm}.}, for which we follow Ref.~\cite{Antusch:2024ypp} (see also  Refs.~\cite{Blanco-Pillado:2013qja,Blanco-Pillado:2017oxo,Cui:2017ufi,Cui:2018rwi,Auclair:2019wcv,Gouttenoire:2019kij,Blasi:2020wpy,Dunsky:2021tih,Buchmuller:2021mbb}).

The gravitational wave spectrum is given by
\begin{align}
&\Omega_\mathrm{GW}(f,t)=\frac{8\pi (G\mu)^2}{3H^2(t)} f \times \label{eq:OGW}
\\&
\sum_{n=1}^\infty  P_n \underbrace{\bigg\{ \frac{2n}{f^2}\int_{z(t)}^{z_c}\frac{dz}{H(z)(1+z)^6}\mathbf{n}\Big(\frac{2n}{f(1+z)},t(z)\Big) \bigg\} }_{C_n}. \nonumber
\end{align}
Here, $P_n$ is the radiation power spectrum of each loop (which we take from Ref.~\cite{Blanco-Pillado:2017oxo}), $\mathbf{n}$ is the loop number density, and $G\mu^2\sum_{n=1}^\infty C_n P_n\equiv \rho_\mathrm{GW}(t,f)$ is the energy density in gravitational waves per unit frequency. Moreover, $z$  represents the redshift (redshift today will be denoted by $z_0$) and the   cosmological dependence is encoded in the Hubble rate, defined as, 
\begin{align}
H(z)=H_0\left( \Omega_\Lambda + (1+z)^3\Omega_\mathrm{mat} + (1+z)^4 \mathcal{G}(z) \Omega_\mathrm{rad} \right)^{1/2},\label{eq:HCDM}
\end{align}
with $H_0=67.8$ km/s/Mpc, $\Omega_\Lambda=1-\Omega_\mathrm{mat}-\Omega_\mathrm{rad}$, $\Omega_\mathrm{mat}=0.308$, and $\Omega_\mathrm{rad}=9.1476\times 10^{-5}$~\cite{Planck:2018vyg}.

The factor that reflects how the number of relativistic degrees of freedom evolves is expressed as 
\begin{align}
\mathcal{G}(z)= \frac{g_\ast(z)g^{4/3}_\mathrm{S}(z_0)}{g_\ast(z_0)g^{4/3}_\mathrm{S}(z)},\label{eq:Gdof}
\end{align}
where $g_\ast(z)$ is the effective number of degrees of freedom and $g_\mathrm{S}(z)$ is the effective number of entropic degrees of freedom at redshift $z$. In the regime with $z\leq10^6$, $\mathcal G(z)=1$. On the other hand, for larger redshift, namely  $z>10^6$, the universe is radiation dominated, and the Hubble rate can be approximated as $H(z)=H_0\Omega_\text{rad}^{1/2}\mathcal G(z)^{1/2}(z+1)^2$. In this regime, the $\mathcal{G}(z)$ factor in Eq.~\eqref{eq:Gdof} includes not only the annihilation of relativistic species in the early universe from the SM, but also the additional degrees of freedom from the MSSM. Therefore, we have 
\begin{align}
&g^\mathrm{MSSM}_*(T)=g_*^\mathrm{SM}(T)  \nonumber
\\& \hspace{30pt}
+\frac {15}{\pi^4}\bigg(32\ J_{+}\big(m_\mathrm{S}/T\big)+94\ J_{-}\big(m_\mathrm{S}/T\big)\bigg), 
\\
&J_\pm(x)=\int_0^\infty\ dy \frac{y^2\sqrt{y^2+x^2}}{\exp\big(\sqrt{y^2+x^2}\big)\pm1},
\end{align}
where, for simplicity, we have assumed all superpartner particles to have a common mass $m_\mathrm{S}=M_\mathrm{SUSY}$  (with a corresponding temperature of $T_\mathrm{S}$). At some early redshift $z^\prime$, for which $\mathcal G(z^\prime)=1$, $T^\prime$ is computed from  
\begin{equation}
 3 m^2_\mathrm{Pl}H^2(z)=\rho(z)=\frac{\pi^2}{30}g_*(T)T^4.
\end{equation}
By utilizing $T^\prime$, $z^\prime$, and imposing entropy conservation we then determine the  temperature as a function of the redshift,   $T(z)$.

To account for an intermediate phase of matter domination, we assume that the associated energy density is transferred to the radiation bath via decay. This process reheats the subdominant radiation component, thereby initiating a second phase of radiation domination. This matter domination phase is parameterized in terms of the dilution factor $D$ and the redshift at its end, $z_E$.  During this matter domination period, $z_E\leq z\lesssim z_E D$, the Hubble rate is given by $H(z)=H_\mathrm{MD}(z+1)^{3/2}$, where $H_\mathrm{MD}$ is fixed by requiring $H(z)$ to be continuous at $z_E$.

Finally,  the loop number density, $\mathbf{n}(\ell,t)$,   is obtained by solving the corresponding partial differential equation~\cite{Buchmuller:2021mbb} ($\ell$ denotes the length of the string).  For $t<t_d= \Gamma_d^{-1/2}$,
\begin{align}
\mathbf{n}(\ell,t)= \mathbf{n}_{S}(\ell,t),    
\end{align}
whereas for $t>t_d$, 
\begin{align}
&\mathbf{n}(\ell,t)=    \mathbf{n}_S\big(\ell+\Gamma G\mu(t-t_d),t_d\big) \times
\\&
\exp\bigg[-\Gamma_d\Big(\ell(t-t_d)+\frac 12G\mu\Gamma(t-t_d)^2\Big)\bigg]\Big(\frac{a(t_d)}{a(t)}\Big)^3, \nonumber
\end{align}
with 
\begin{equation}
\mathbf{n}_S(\ell,t)=\int^t_{t_c}d\tau\bigg(\frac{a(\tau)}{a(t)}\bigg)^3S(\ell+\Gamma \,G\mu(t-\tau),\tau).
\end{equation}
The form of the  function
$S$ is obtained from simulations~\cite{Blanco-Pillado:2013qja}, which for matter domination era has the following form:
\begin{align}
S(\ell,t)\approx   & \; 
d^{-5}_H(t)\frac{5.34}{(\ell/d_H(t))^{1.69}}\Theta\big(0.06-\ell/d_H(t)\big)
\nonumber\\& \times
\Theta\big(\ell/d_H(t)-\Gamma \,G\mu\big).
\end{align}
The second $\Theta$ function is introduced to cut off the $\ell/d_H(t)\rightarrow0$ divergence~\cite{Blanco-Pillado:2013qja}. Here, $d_H$ is the Hubble radius, which can be determined following Ref.~\cite{Antusch:2024ypp}. And for radiation dominated era it is given by
\begin{align}
  S(\ell,t)\approx   d^{-5}_H(t) 92.13\,\delta\big(\ell/d_H(t)-0.05\big).
\end{align}
Using the recipes given above, we compute the gravitational wave spectrum for a metastable cosmic string network.

\bibliographystyle{style}
\bibliography{reference}
\end{document}